\documentclass[aps,prl,twocolumn,superscriptaddress,nofootinbib]{revtex4-1}

\usepackage[utf8]{inputenc}
\usepackage{epsfig} 
\usepackage[colorlinks,breaklinks]{hyperref}
\usepackage{nicefrac}
\usepackage{bm}
\usepackage{dcolumn}
\usepackage{graphics}
\usepackage{amsmath}
\usepackage{amssymb}
\usepackage{isotope}
\usepackage{xspace}
\usepackage{suffix}

\newcommand{\bra}{\langle}
\newcommand{\ket}{\rangle}

\WithSuffix\newcommand\ket*[1]{|#1\rangle}
\WithSuffix\newcommand\bra*[1]{|#1\rangle}

\newcommand{\be}{\begin{equation}}
\newcommand{\ee}{\end{equation}}
\newcommand{\bea}{\begin{align}}
\newcommand{\eea}{\end{align}}

\newcommand{\evec}{\ensuremath{\boldsymbol{\eta}}}

\newcommand{\apriori}{\textit{a priori}\xspace}
\WithSuffix\newcommand\apriori*{\textit{a-priori}\xspace}

\newcommand{\ii}{\mathrm{i}}

\newcommand{\LO}{\text{LO}}
\newcommand{\NLO}{\text{NLO}}

\begin{document}

\title{Four-Body Scale in Universal Few-Boson Systems}

\author{B.~Bazak}
\affiliation{The Racah Institute of Physics, The Hebrew University, 9190401, 
Jerusalem, Israel}

\author{J.~Kirscher}
\affiliation{Department of Physics, The City College of New York, New York,
  New York 10031, USA}
\affiliation{Theoretical Physics Division, School of Physics and Astronomy,
The University of Manchester, Manchester, M13 9PL, United Kingdom}

\author{S.~König}
\affiliation{Institut für Kernphysik, Technische Universität Darmstadt,
64289 Darmstadt, Germany}
\affiliation{ExtreMe Matter Institute EMMI,
GSI Helmholtzzentrum für Schwerionenforschung GmbH,
64291 Darmstadt, Germany}

\author{M.~Pav\'on Valderrama}
\affiliation{School of Physics and Nuclear Energy Engineering,
International Research Center for Nuclei and Particles in the Cosmos and
Beijing Key Laboratory of Advanced Nuclear Materials and Physics,
Beihang University, Beijing 100191, China}

\author{N.~Barnea}
\affiliation{The Racah Institute of Physics, The Hebrew University, 9190401, 
Jerusalem, Israel}

\author{U.~van Kolck}
\affiliation{Institut de Physique Nucléaire, CNRS-IN2P3, 
Université Paris-Sud, Université Paris-Saclay, 91406 Orsay, France}
\affiliation{Department of Physics, University of Arizona,
Tucson, Arizona 85721, USA}

\date{\today}

\begin{abstract}
The role of an intrinsic four-body scale in universal few-boson systems is the
subject of active debate.
We study these systems within the framework of effective field theory.
For systems of up to six bosons we establish that no four-body 
scale appears at leading order (LO).  
However, we find that at next-to-leading (NLO) order a four-body force is
needed to obtain renormalized results for binding energies.
With the associated parameter fixed to the binding energy of the four-boson 
system, this force is shown to renormalize the five- and six-body systems as 
well.
We present an original ansatz for the short-distance limit of the 
bosonic $A$-body wave function from which we conjecture that new $A$-body
scales appear at N$^{A-3}$LO.
As a specific example, calculations are presented for clusters of helium atoms. 
Our results apply more generally to other few-body systems governed by 
a large scattering length, such as light nuclei and halo states,
the low-energy properties of which 
are independent of the detailed internal structure of the constituents.
\end{abstract}

\maketitle

The universal aspects of few-body systems with large scattering length have
attracted a lot of attention in recent years~\cite{BraHam06,GreGiaPer17}, 
largely owing to well-controlled experiments involving ultracold atomic gases 
where the scattering length can be tuned arbitrarily via Feshbach 
resonances~\cite{ChiGriJul10}.
Nuclear physics, where the scattering lengths in both nucleon-nucleon $S$-wave
channels are significantly larger in magnitude than the interaction range set by
the pion mass, falls into the same universality class.
Another interesting example is given by atomic \isotope[4]{He} clusters, where
the two-body scattering length also happens to be much larger than the van der
Waals length. 

These systems share a pronounced separation of scales. When the 
scattering length $a$ significantly exceeds the force range,
the system properties become independent of the force details, 
which can be represented by contact interactions, 
in analogy with the multipole expansion of classical electrodynamics.
Effective field theory (EFT) implements this idea systematically starting 
from a leading order (LO) with only Dirac delta functions.
Finite-range corrections are accounted for at higher orders through delta
functions with derivatives.
In the two-body sector, this expansion around the zero-range limit is 
equivalent~\cite{vanKolck:1998bw} to Fermi's pseudopotential~\cite{Fer36},
to nontrivial boundary conditions ~\cite{BetPei35},
and to the effective range expansion~\cite{Bet49}.

This idea extends to $A$-body systems: $A$-body forces,
which capture aspects of the underlying potential that only manifest
themselves for $A>2$,
are ordered according to their relevance to low-energy physics.
For three identical particles with spin statistics not precluding a 
totally symmetric spatial wave function---bosons or nucleons, for instance---the
zero-range limit is not well defined with only two-body interactions
due to bound-state collapse~\cite{Thomas:1935zz}. 
In the EFT framework, despite expectations based on dimensional analysis,
a three-body contact interaction must enter at LO to ensure 
renormalization-group (RG) invariance~\cite{BedHamKol99,BedHamKol99b}.
It introduces a scale that in the unitary limit ($a\to \infty$) determines 
the position of a geometric tower of three-body bound 
states~\cite{Efimov:1970zz}.
These Efimov trimers have been observed in cold-atom systems~\cite{Kra06,Hua14}.

It is of fundamental interest to understand whether this phenomenon repeats 
in larger systems: when does an additional particle bring in a new scale
from a higher-body contact interaction?
Once such a scale appears, universality is reduced and the properties 
of the corresponding system can no longer be predicted entirely on the basis 
of systems with fewer particles.

The importance of a four-body parameter has in fact been the subject of
active debate in the literature~\cite{PlaHamMei04,Platter05,HanBlu06,%
YamTomDel06,HamPla07,SteDinGre09,HadYamTom11}.
In contrast to a zero-range model~\cite{YamTomDel06,HadYamTom11}, early EFT
studies~\cite{PlaHamMei04,Platter05,HamPla07} of the four-body system found that
no four-body force (and thus no four-body scale) is required at LO.
Recently, Ref.~\cite{BazEliKol16} also established the absence of LO 
higher-body forces for systems of up to six particles.
Lack of further LO scales means that Efimov towers exist for more than
three bosons~\cite{HamPla07,SteDinGre09,vonStecher:2010,Gattobigio:2011ey,%
vonStecher:2011zz,Gattobigio:2012tk} and that there are correlations between
cluster and trimer binding energies, the atomic equivalent of the nuclear Tjon
line~\cite{Tjo75} for four-boson clusters~\cite{PlaHamMei04} and its
generalization for five- and six-boson clusters~\cite{BazEliKol16}.
The properties of unitary bosonic matter are universal when written in terms of
a three-body energy~\cite{Carlson:2017txq}, a property that might be testable in
cold-atom experiments~\cite{Makotyn:2014,Fletcher:2017}.

Here, we go beyond LO and address the question whether the naïvely construed
next-to-leading-order (NLO) part of the  EFT expansion, where range corrections
enter in the form of contact interactions with derivatives, is properly
renormalized.
While subleading orders have been included perturbatively in the three-boson
system with success~\cite{Ji:2011qg,Ji:2012nj}, our work is the first to
examine more-boson systems in this way.
Our central result is that a four-body force is required at NLO, which, once
fixed to a single four-body observable, suffices to stabilize clusters of up to
at least six bosons.
The relatively small resulting changes at NLO bode well for the convergence of
the EFT expansion.

\paragraph{EFT description.}

A system of nonrelativistic spinless bosons of mass $m$ interacting via a 
short-range force can be described by the Lagrangian density
\begin{equation}
 \mathcal L = \psi^\dagger \left(\ii\partial_0 
  + \frac{\nabla^2}{2m} \right) \psi
  - \frac{C_0^{(0)}}{2} (\psi^\dagger \psi)^2 
  - \frac{D_0^{(0)}}{6}  (\psi^\dagger \psi)^3 + \cdots \,,
\label{eq:Leff}
\end{equation}
where $\psi$ is the field operator, $C_0^{(0)}$ and $D_0^{(0)}$ are 
low-energy constants (LECs), and the ellipsis represents terms with more 
fields and/or more derivatives, entering at higher orders.
The LECs' super- and subscripts denote, respectively, the order in the EFT
expansion and the powers of momenta involved.

Translated to the language of ordinary quantum mechanics, the interaction terms
in Eq.~\eqref{eq:Leff} give rise to delta-function potentials, which need to 
be regularized.
We choose here a separable form, 
$ V_2^{(0)} = C_0^{(0)}|g\ket\bra g|$, 
where $g$ represents a Gaussian regulator in momentum
space, $ \bra {\bf q} | g \ket = \exp(-q^2/\Lambda^2) \equiv g(q^2)$.
In coordinate space, this corresponds to a smeared-out delta function which
tends to a delta function as the cutoff parameter $\Lambda\to \infty$.
Observables must not depend on the arbitrary regularization except for terms 
that decrease as $\Lambda$ increases.
This is achieved \textit{via} renormalization, when the LEC ``runs'' with the
cutoff, $C_0^{(0)} = C_0^{(0)}(\Lambda)$, in such a way that a chosen
observable---for example, the scattering length---remains fixed to its physical 
value.

The term involving $D_0^{(0)}$ parametrizes the three-body force at LO.
We include it in the form $V_3^{(0)} = D_0^{(0)}|\xi\ket\bra \xi|$, where
$\bra {\bf q}_1 {\bf q}_2 |\xi\ket = g(q_1^2+3q_2^2/4)$ regulates the three-body
system described by the Jacobi momenta $q_{i}$.
Renormalization is achieved when one three-body observable---for example 
a trimer energy---is kept fixed.
$D_0^{(0)}(\Lambda)$ has a log-periodic form~\cite{BedHamKol99,BedHamKol99b}
representing an RG limit cycle.

Range corrections enter at NLO in the form of a term that involves four fields 
and two derivatives, with a new LEC $C^{(1)}_2$ to be determined from a second 
two-body observable. The corresponding potential can be written in momentum 
space as
\begin{equation}
 \bra {\bf k}|V^{(1)}_{2}|{\bf k}'\ket = C^{(1)}_2 g(k^2)
 \left(k^2+k'^2\right) g(k'^2) \,.
\end{equation}
There are also corrections to the LO LECs that do not introduce new parameters
because they merely ensure that the renormalization conditions used at LO remain
satisfied at NLO.  
Although the LO interactions must be treated nonperturbatively, NLO consists
of a single insertion of the NLO potential.
Renormalization cannot be achieved for positive effective range, as is the
case here, when an inconsistent subset of higher-order 
corrections is included by the nonperturbative solution of the Schr\"odinger 
equation with the NLO potential~\cite{Beane:1997pk}.

\paragraph{Numerical methods.}

We employ two independent numerical methods to calculate the $A$-boson binding 
energies, both treating NLO corrections perturbatively.

In the first approach, which is more efficient for a precise numerical 
determination of $D_0^{(0)}(\Lambda)$, we calculate $A=3,4$ binding energies by
solving, respectively, the Faddeev and Faddeev-Yakubovsky (FY) equations.
We employ the same numerical framework as in Ref.~\cite{KonGriHam16}, 
which is an implementation of the formalism discussed in
Refs.~\cite{Kamada92,PlaHamMei04,Platter05}.
The central idea is to decompose the full wave functions into 
Faddeev(-Yakubovsky) components which are related by Bose symmetry.
For $A=3$, we express the wave function
$\ket*{\Psi_{3}^{(0)}} = (1 + P) \ket*{\psi}
+ \ket*{\psi_3}$ in terms of the Faddeev components $\ket*{\psi}$ and
$\ket*{\psi_3}$, where $1 + P$ with $P = P_{12}P_{23} + P_{13}P_{23}$ is an
operator that enforces Bose symmetry through a combination of appropriate
permutations $P_{ij}$ of the individual particles.
One obtains the system of equations
\begin{align}
 \ket*{\psi} &= G_0\,t\,P \ket*{\psi} + G_0\,t\,\ket*{\psi_3} \,,
 \nonumber \\
 \ket*{\psi_3} &= G_0\,t_3\,(1+P) \ket*{\psi} \,,
\label{eq:Faddeev-3B}
\end{align}
where $G_0$ denotes the free three-body Green's function and the operators $t$ 
and $t_3$ are solutions of Lippmann-Schwinger equations with, respectively, 
$V_2^{(0)}$ and $V_3^{(0)}$ as driving terms.
It is an advantage of the separable regulator we use that these
operators can be derived analytically in closed form.
The solution for Eq.~\eqref{eq:Faddeev-3B} is obtained in momentum space by
projection onto partial-wave states $\ket*{q_1 q_2,l_1 l_2}$, where
$l_{1,2}$ are orbital angular-momentum quantum numbers corresponding to the
Jacobi momenta $q_{1,2}$; they are coupled to total angular momentum zero for 
the states we consider in this letter.
Upon discretization on a momentum grid, Eq.~\eqref{eq:Faddeev-3B} yields a 
homogeneous matrix equation that depends on energy via $G_0$ and $t$.
Bound states are found at those energies where the matrix has a unit 
eigenvalue.
The wave function components are obtained by solving the corresponding 
homogeneous equations.
Similarly, for $A=4$, $\ket*{\Psi_{4}^{(0)}} = (1 + P_{34} +
PP_{34})(1+P)\ket*{\psi_A} + (1 + P)(1 + \tilde{P})\ket*{\psi_B}$ involves the
additional permutation operator $\tilde{P} \equiv P_{13}P_{24}$ as well as
components $\ket*{\psi_A}$ and $\ket*{\psi_B}$ that correspond 
to partitions into, respectively, $3+1$ and $2+2$ clusters.

In the second approach, which is more efficient for systems with more particles,
we expand the coordinate-space wave function in a correlated Gaussian 
basis~\cite{SuzVar98},
\begin{equation}
 \Psi_{A}^{(0)}(\evec)=
 \sum_i c_i \, \hat {\mathcal S} \,
 \exp\left({-}\frac{1}{2}\evec^T\!A_i \evec\right) \,,
\end{equation}
where $\evec$ collects the $A-1$ Jacobi vectors $\evec_j$, $A_i $ is an
$(A-1)\times(A-1)$ real, symmetric, and positive-definite matrix, and
$\hat{\mathcal S}$ is a symmetrization operator. 
The coefficients $\{c_i\}$ and the energy are determined by solving a
generalized eigenvalue problem.  An important feature of the Gaussian basis is
that it can deal with both short ($\sim\Lambda^{-1}$) and long  
($\sim a$) length scales.
To optimize our basis we use the stochastic variational method
(SVM)~\cite{SuzVar98}, where the elements of the matrix $A_i$ are chosen
randomly taking at each step the element that gives the lowest energy.
By the variational principle, the method is guaranteed to give upper bounds for
the binding energies.
The implementation of this method here follows Ref.~\cite{BazEliKol16}.

Our choice of a separable Gaussian regulator significantly simplifies
the Faddeev equations.
With SVM we could verify that our results are reproduced with a 
nonseparable regulator
made of local Gaussians in configuration space.

\paragraph{Results.}

While our conclusions are generally valid for other universal systems, such as 
ultracold atomic gases or atomic nuclei, for concreteness we calculate here the
energies of small clusters of $^4\rm{He}$ atoms.
The $^4\rm{He}$ atomic system is characterized by a scattering length
$a\approx90\,\rm{\AA}$, which is much larger than the van der Waals length 
$\approx 5.4\,\rm{\AA}$ that describes the long-range part of the interatomic 
potential.
The dimer was measured experimentally to have a binding energy of about 
$1.5$~mK~\cite{LuoMcBKim93,GriSchToe00,Zel16}.
Two Efimov trimers were measured \cite{SchToe96,Kun15}, which are the 
remains of the otherwise infinite geometric tower of Efimov states 
that emerges as $a\to\infty$.
Larger clusters
are predicted~\cite{BluGre00,GuaKorNav06,HiyKam12a,BazValBar19} 
by modern He-He pair potentials~\cite{JanAzi95,PrzCenKom10}, 
but have not yet been observed.

Three data points are needed to fix the coefficients of our EFT up to NLO,
which we choose as the two-body scattering length and effective range, as well
as the binding energy of the excited trimer.
In order to compare with heavier-cluster predictions, we take the values
calculated from a potential, in particular the modern PCKLJS
potential~\cite{PrzCenKom10,HiyKam12a}.
Once enough data on helium clusters become available, we can 
let go of potential-model input.
For now, we use this two-body potential as a possible
representation of short-distance physics; the inclusion of more complicated 
interactions~\cite{CenPatSza09} would not affect our conclusions.
The dimer binding energy here is $B_2=1.615$~mK, and indeed our EFT converges
well toward this value, with $B_2^{\LO} = 0.918\,B_2$
and $B_2^{\NLO} = 0.991\,B_2$.
We use the Faddeev equations to fix $D_0^{(0)}(\Lambda)$ and then find good
agreement between the two methods for the ground state trimer binding energy,
$B_{3}^{\LO} = 98.1\,B_2$ and $B_{3}^{\NLO} = 73.1\,B_2$, to be
compared with the direct potential-model result
$B_3 = 81.6\,B_2$~\cite{HiyKam12a}.

\begin{figure}
\centering
  \includegraphics[width=\columnwidth]{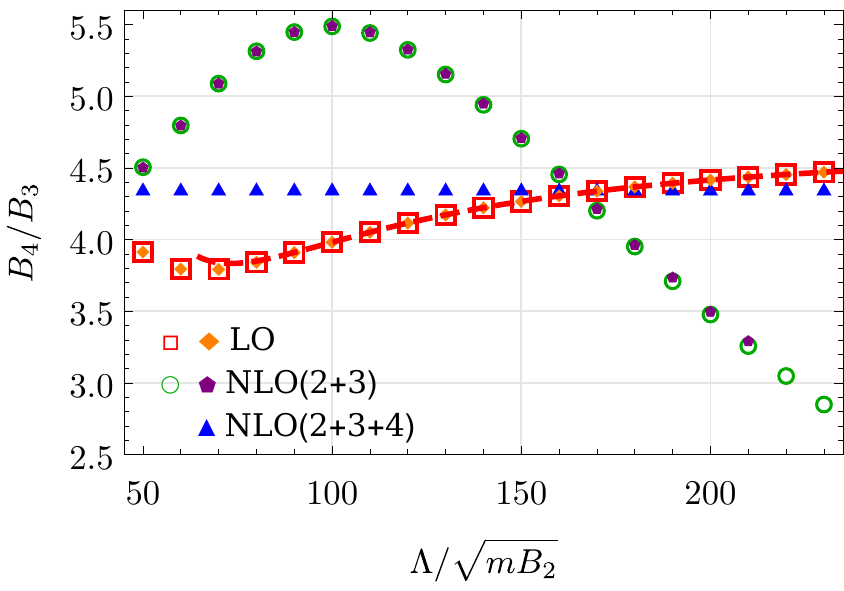}
  \caption{\label{fig:B4} 
  The tetramer binding energy in units of the
  trimer ground-state energy is plotted as function of the cutoff in units of 
  $\sqrt{m B_2}$.
  LO and NLO results without a four-body force from the FY (orange diamonds and
  purple pentagons) and SVM (red squares and green circles) methods are in very
  good agreement.
  They are compared to the result calculated directly~\cite{HiyKam12a} from
  the PCKLJS potential, which coincides (by construction)
  with the NLO result with a four-body force (blue triangles).
  The red dashed curve is a fit in powers of $\Lambda^{-1}$.}
\end{figure}

EFT calculations for four-atom systems so far are only available at LO. 
Here we confirm the pioneering result~\cite{PlaHamMei04}
that the LO tetramer ground-state energy converges as the cutoff $\Lambda$ 
is increased.
We proceed for the first time to NLO, where we observe that, in contrast,
the tetramer energy does \emph{not} converge once range corrections are
included---it instead diverges roughly linearly within the investigated cutoff
range.
Our LO and NLO results for the tetramer ground-state energy as a function of
$\Lambda$ are shown in Fig.~\ref{fig:B4}.
The two methods agree very well.

The observed divergence is a clear indication that a four-body force is 
required at NLO, much earlier than one would expect from a naïve counting of
many-body forces.
This promotion is analogous to that of the three-body force to LO.
The simplest four-body force is a contact interaction without derivatives: 
$V_4^{(1)}=F_0^{(1)}|\zeta\ket\bra\zeta|$, where 
$\bra {\bf q}_1 {\bf q}_2 {\bf q}_3| \zeta \ket = 
g\big(q_1^2+3q_2^2/4+2q_3^2/3\big)$
in the same regularization as before.
The LEC $F_0^{(1)}(\Lambda)$ is determined by demanding that the tetramer energy
is fixed at the value calculated directly from the potential
model~\cite{HiyKam12a}.

\begin{figure}
 \centering
 \includegraphics[width=\columnwidth]{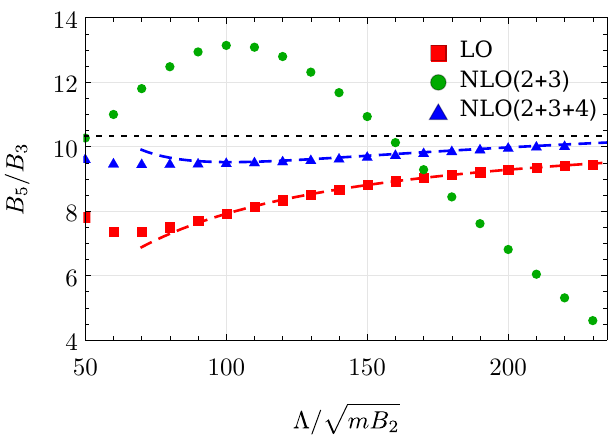}
 \caption{\label{fig:B5} 
  The pentamer ground-state energy is plotted as
  function of the cutoff in the same units as Fig.~\ref{fig:B4}.  
  Shown are results at LO (red squares), NLO without a four-body force (green 
  circles), and NLO with the four-body force that renormalizes the four-body
  system (blue triangles).
  The colored dashed curves are fits in powers of $\Lambda^{-1}$. 
  The result calculated~\cite{BluGre00} from the LM2M2 potential is the dotted
  line.}
\end{figure}

\begin{figure}
 \centering
 \includegraphics[width=\columnwidth]{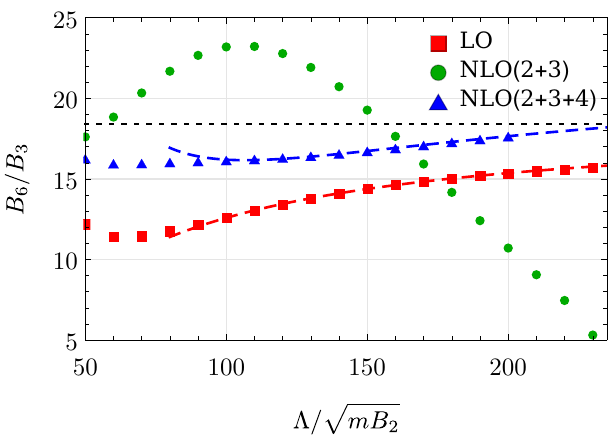}
 \caption{\label{fig:B6} 
  The hexamer ground-state energy is plotted as
  function of the cutoff in the same units as Fig.~\ref{fig:B4}.  
  Symbols are as in Fig.~\ref{fig:B5}.}
\end{figure}

With the NLO four-body force thus determined, one may wonder whether higher-body
forces appear at the same order.
We find that this is not the case when we study the pentamer ground-state energy
up to NLO with SVM, as shown in Fig.~\ref{fig:B5}.
At LO the results converge with $\Lambda$, in agreement with the conclusion of
Ref.~\cite{BazEliKol16} that no five-body term is needed at this order.
Without a four-body force a divergence is observed at NLO, analogously to the
one observed for the tetramer energy, but once the NLO four-body force is
included, we find the five-body system properly renormalized as well.
This adds confidence in our order assignment for the dominant four-body force.
Similar conclusions hold for the six-atom system, as can be seen from the
hexamer ground-state energy in Fig.~\ref{fig:B6}.
Although the SVM calculation becomes more difficult as the number of particles
increases and the results are therefore less conclusive for $A=6$, we see
overall the same pattern as before.
There is no need for a six-body force up to NLO, either.

Numerical calculations are limited to finite cutoffs.
In a renormalized theory, the residual cutoff dependence
can be absorbed in higher-order operators, which scale as inverse
powers of the breakdown scale.
Asymptotic ($\Lambda\to \infty$) values obtained from fitting
the numerical results with polynomials in $\Lambda^{-1}$
are given in Table~\ref{tbl:HeBE}, where the reported error is
that from the extrapolation alone.
A reasonable estimate of the EFT truncation error at NLO is the square of the
relative change from LO to NLO.
No results based on the PCKLJS potential are available to compare with;
however, since the tetramer energies based on the PCKLJS and LM2M2 differ
by only 2\%, we list the latter~\cite{BluGre00} as representative results.

\begin{table}
 \centering
 \renewcommand{\arraystretch}{1.25}%
 \begin{tabular}
 {c@{\hspace{3mm}}|c@{\hspace{3mm}} c@{\hspace{3mm}} 
 c@{\hspace{3mm}}c@{\hspace{3mm}} c}
 \hline
 \hline 
 & LO & NLO & PCKLJS & LM2M2
 \\ 
 \hline
 $B_4/B_3$ & 4.8(1)  & 4.35$(*)$ & 4.35 & 4.44(1)   \\
 $B_5/B_3$ & 10.8(5) & 11.3(3)& ---  & 10.33(1)  \\
 $B_6/B_3$ & 18(2)   & 22(3)  & ---  & 18.41(2)  \\
 \hline
 \hline
 \end{tabular}
 \caption{\label{tbl:HeBE} The $A$-body $^4\rm{He}$ binding energies, in units 
  of the trimer binding energy, for $A=4,5,6$. 
  $(*)$ indicates a fit value.
  Our results are compared to those 
  obtained with the PCKLJS~\cite{HiyKam12a} and LM2M2~\cite{BluGre00} 
  potentials.}
\end{table}

Our numerical calculations are limited by a cutoff value above which
an unphysical, deep trimer state appears~\cite{BedHamKol99,BedHamKol99b}.
However, considering an appropriate ansatz for the wave function, the 
need for a four-body force at NLO can also be derived analytically.
By using the RG framework of Ref.~\cite{ValPhi15}, we can connect the counting
of the $A$-body LECs with the power-law behavior of the $A$-body wave function
at short distances.
Our ansatz for the latter is $\Psi_A^{(0)} \to 
\phi_{A}/(\prod_j |\eta_j|)$ with 
$\phi_{A}$ a function that does not exhibit power-law behavior for
$\eta_j\to0$.
This ansatz is derived from the Schrödinger equation with short-range
forces in the unitary limit:
the $1/|\eta_j|$ factors are a trivial consequence of the absence of long-range
interactions, while $\phi_{A}$ is a non-trivial consequence of the
(complicated) boundary conditions induced by the short-range forces.
The exact form of $\phi_{A}$ is not known in general, but it is
irrelevant for power counting.
This counting follows from considering matrix elements of $A$-body contact 
potentials between the wave functions, the short-distance behavior of which
determines at which order they are required.
For $A=2,3$ our conjecture is readily verified analytically.
Our numerically obtained $A=3,4$ wave functions approximate this behavior for
distances shorter than the trimer or tetramer sizes and larger than
$\Lambda^{-1}$, i.e., for the expected domain of validity of the ansatz, which
we can check by assuming a log-periodic form for $\phi_{A}$.
Further trust in this ansatz stems from its correct prediction of LO 
two- and three-body forces and the absence of a four-body force at LO.
The validity of the ansatz for $A \geq 5$ is a conjecture, 
which implies even less significance for all higher-body forces,
in agreement with our numerical results.
As such, it establishes the power counting of all $A$-body forces for $A$-boson
systems: new scales appear at N$^{A-3}$LO.

\paragraph{Conclusions.} 

We find a large dependence of the ground-state energies for $A=4,5,6$ bosons on
the regulator when NLO two-body range corrections are added perturbatively 
to LO.
A four-body force is necessary and sufficient for renormalization at this order.
For $A= 4$ this result applies also to fermions with four internal states,
such as the nucleon.
Previous calculations for the $^4\rm{He}$ nucleus
~\cite{Kirscher:2009aj,Lensky:2016djr,Bansal:2017pwn} could 
not observe this effect because range corrections were treated 
nonperturbatively, thereby 
breaking RG invariance already at the two-body level.
It will be interesting to investigate in future work to what extent
the enhancement of many-body forces discussed here is modified in nuclear
systems with $A>4$ due to the Pauli principle.

The low order of the dominant four-body force offers an explanation
for the controversy on the importance of a four-body scale~\cite{PlaHamMei04,%
Platter05,HanBlu06,YamTomDel06,HamPla07,SteDinGre09,HadYamTom11}, 
on one hand, and the need for an additional parameter in the description
of $^4\rm{He}$ droplets~\cite{Kievsky:2017mjq}, on the other.
The absence of higher-body forces up to NLO ensures that correlations between
higher-body and four-body energies survive at this order.
The relatively small size of the full NLO corrections suggests that the EFT
expansion is working well for the $^4\rm{He}$ clusters considered here, 
despite their large binding compared with the trimer.
We plan to extend our calculations to light nuclei in the near future.

\begin{acknowledgments}
We thank Lorenzo Contessi and Francesco Pederiva for useful discussions.  
This letter was supported in part by 
the U.S. NSF under Grants No. PHY15-15738 (J.K.) and No. PHY--1614460 (S.K.); 
the U.S. DOE, NUCLEI SciDAC Collaboration Award No. DE-SC0008533 (S.K.)
and Office of Science, Office of Nuclear Physics, under Award
No. DE-FG02-04ER41338 (U.v.K.);
the ERC Grant No. 307986 STRONGINT (S.K.);
the Deutsche Forschungsgemeinschaft (German Research Foundation) -- 
Projektnummer 279384907 -- SFB 1245 (S.K.);
the National Natural Science Foundation of China under Grants No. 11522539 
and No. 11735003 (M.P.V.);
the Fundamental Research Funds for the Central Universities (M.P.V.);
the Thousand Talents Plan for Young Professionals (M.P.V.); 
the Israel Science Foundation under Grant No. 1308/16 (N.B.);
and 
the European Union Research and Innovation program 
Horizon 2020 under Grant No. 654002 (U.v.K.).
Some of the numerical computations were performed on the Lichtenberg
high-performance computer of the TU Darmstadt as well as at the 
J\"ulich Supercomputing Center.
\end{acknowledgments}



\begin{thebibliography}{99}

\bibitem{BraHam06}
  E.~Braaten and H.-W.~Hammer,
  Universality in few-body systems with large scattering length,
  Phys.\ Rep.\ {\bf 428}, 259 (2006).

\bibitem{GreGiaPer17}
  C. H. Greene, P. Giannakeas, and J. Pérez-Ríos, 
  Universal few-body physics and cluster formation,
  Rev. Mod. Phys. {\bf 89}, 035006 (2017).

\bibitem{ChiGriJul10}
  C.~Chin, R.~Grimm, P.~Julienne, and E.~Tiesinga,
  Feshbach resonances in ultracold gases,
  Rev.\ Mod.\ Phys.\ {\bf 82}, 1225 (2010).

\bibitem{vanKolck:1998bw}
  U.~van Kolck,
  Effective field theory of short-range forces,
  Nucl. Phys. {\bf A645}, 273 (1999).

\bibitem{Fer36}
  E.~Fermi,
  Motions of neutrons in hydrogenous substances,
  Ric. Sci. {\bf 7}, 13 (1936).

\bibitem{BetPei35}
  H.~A.~Bethe, and R.~Peierls,
  Quantum theory of the diplon,
  Proc.\ R.\ Soc.\ A {\bf 148}, 146 (1935).
  
\bibitem{Bet49}
  H.~A.~Bethe,
  Theory of the effective range in nuclear scattering,
  Phys.\ Rev.\ {\bf 76}, 38 (1949).
  
\bibitem{Thomas:1935zz}
  L.~H.~Thomas,
  The interaction between a neutron and a proton and the structure of 
  H$^3$,
  Phys.\ Rev.\ {\bf 47}, 903 (1935).

\bibitem{BedHamKol99}
  P.~F.~Bedaque, H.-W.~Hammer, and U.~van~Kolck,
  Renormalization of the three-body system with short-range
  interactions,
  Phys.\ Rev.\ Lett.\ {\bf 82}, 463 (1999).

\bibitem{BedHamKol99b}
  P.~F.~Bedaque, H.-W.~Hammer, and U.~van~Kolck,
  The three-boson system with short-range interactions,
  Nucl.\ Phys.\ {\bf A646}, 444 (1999).

\bibitem{Efimov:1970zz}
  V.~Efimov,
  Energy levels arising form the resonant two-body forces in a 
  three-body system,
  Phys.\ Lett.\ {\bf 33B}, 563 (1970).

\bibitem{Kra06}
  T. Kraemer~\textit{et al.},
  Evidence for Efimov quantum states in an ultracold gas of
  caesium atoms,
  Nature (London) {\bf 440}, 315 (2006).

\bibitem{Hua14}
  Bo Huang~\textit{et al.},
  Observation of the second triatomic resonance in Efimov's scenario,
  Phys.\ Rev.\ Lett.\ {\bf 112}, 190401 (2014). 

\bibitem{PlaHamMei04}
  L.~Platter, H.-W.~Hammer, and U.-G.~Mei{\ss}ner,
  Four-boson system with short-range interactions,
  Phys.\ Rev.\ A {\bf 70}, 052101 (2004).

\bibitem{Platter05}
  L.~Platter,
  From Cold Atoms to Light Nuclei: The Four-Body Problem in an 
  Effective Theory with Contact Interactions,
  Doctoral thesis (Dissertation), University of Bonn, 2005.

\bibitem{HanBlu06}
  G.~J.~Hanna and D.~Blume
  Energetics and structural properties of
  three-dimensional bosonic clusters near threshold,
  Phys.\ Rev.\ A {\bf 74}, 063604 (2006).

\bibitem{YamTomDel06} 
  M.~T. Yamashita, L.~Tomio, A.~Delfino, and T.~Frederico,
  Four-boson scale near a Feshbach resonance,
  Europhys.\ Lett.\ {\bf 75}, 555 (2006)

\bibitem{HamPla07}
  H.-W.~Hammer and L.~Platter,
  Universal properties of the four-body system
  with large scattering length,
  Eur.\ Phys.\ J.\ A {\bf 32}, 113 (2007).

\bibitem{SteDinGre09}
  J.~von Stecher, J.~P.~D'Incao, and C.~H.~Greene,
  Signatures of universal four-body phenomena and
  their relation to the Efimov effect,
  Nat.\ Phys.\ {\bf 5}, 417 (2009).

\bibitem{HadYamTom11} 
  M.~R.~Hadizadeh, M.~T.~Yamashita, L.~Tomio, A.~Delfino, and T.~Frederico,
  Scaling Properties of Universal Tetramers,
  Phys.\ Rev.\ Lett.\ {\bf 107}, 135304 (2011).

\bibitem{BazEliKol16}
  B.~Bazak, M.~Eliyahu, and U.~van~Kolck,
  Effective field theory for few-boson systems,
  Phys.\ Rev.\ A {\bf 94}, 052502 (2016).

\bibitem{vonStecher:2011zz}
  J. von Stecher,
  Five- and Six-Body resonances Tied to an Efimov Trimer,
  Phys. Rev. Lett. {\bf 107}, 200402 (2011).

\bibitem{Gattobigio:2011ey}
  M. Gattobigio, A. Kievsky, and M. Viviani,
  Spectra of helium clusters with up to six atoms using soft-core 
  potentials,
  Phys.\ Rev.\ A {\bf 84}, 052503 (2011).

\bibitem{vonStecher:2010}
  J. von Stecher,
  Universal bound cluster states of Efimov character,
  J. Phys. B {\bf 43}, 101002 (2010).

\bibitem{Gattobigio:2012tk} 
  M.~Gattobigio, A.~Kievsky, and M.~Viviani,
  Energy spectra of small bosonic clusters having a large two-body 
  scattering length,
  Phys.\ Rev.\ A {\bf 86}, 042513 (2012).

\bibitem{Tjo75}
  J.~A.~Tjon,
  Bound states of $^4$He with local interactions,
  Phys.\ Lett.\ {\bf 56B}, 217 (1975).

\bibitem{Carlson:2017txq} 
  J.~Carlson, S.~Gandolfi, U.~van Kolck, and S.~A.~Vitiello,
  Ground-state properties of unitary bosons: from clusters to matter,
  Phys.\ Rev.\ Lett.\ {\bf 119}, 223002 (2017).

\bibitem{Makotyn:2014}
  P. Makotyn {\it et al.},
  Universal dynamics of a degenerate unitary Bose gas,
  Nat.\ Phys.\ {\bf 10}, 116 (2014).

\bibitem{Fletcher:2017}
  R. J. Fletcher {\it et al.},
  Two and Three-body Contacts in the Unitary Bose Gas,
  Science {\bf 355}, 377 (2017).

\bibitem{Ji:2011qg} 
  C.~Ji, D.~R.~Phillips, and L.~Platter,
  The three-boson system at next-to-leading order in an effective field
  theory for systems with a large scattering length,
  Ann.\ Phys.\ {\bf 327}, 1803 (2012).

\bibitem{Ji:2012nj}
  C. Ji and D. R. Phillips,
  Effective Field Theory Analysis of Three-Boson Systems at
  Next-To-Next-To-Leading Order,
  Few-Body Syst.\ {\bf 54}, 2317 (2013).

\bibitem{Beane:1997pk}
  S. R.~Beane, T. D.~Cohen, and D. R.~Phillips,
  The Potential of effective field theory in NN scattering,
  Nucl.\ Phys.\ {\bf A632}, 445 (1998).

\bibitem{KonGriHam16}
  S.~König, H.~W.~Grießhammer, H.-W.~Hammer, and U.~van~Kolck,
  Nuclear Physics Around the Unitarity Limit,
  Phys.\ Rev.\ Lett.\ {\bf 118}, 202501 (2017).

\bibitem{Kamada92}
  H.~Kamada and W.~Glöckle,
  Solutions of the Yakubovsky equations for four-body model systems,
  Nucl.\ Phys.\ {\bf A548}, 205 (1992).

\bibitem{SuzVar98}
  Y.~Suzuki and K.~Varga,
  Stochastic Variational Approach to Quantum-Mechanical Few-Body
  Problems,
  (Springer,Berlin Heidelberg, 1998.)

\bibitem{LuoMcBKim93}
  F.~Luo, G.~C.~McBane, G.~Kim, C.~F.~Giese, and W.~R.~Gentry,
  The weakest bond: Experimental observation of helium dimer,
  J.\ Chem.\ Phys.\ {\bf 98}, 3564 (1993).

\bibitem{GriSchToe00}
  R.~E.~Grisenti, W.~Sch\"ollkopf, J.~P.~Toennies, G.~C.~Hegerfeldt, T.~Kohler, 
  and M.~Stoll,
  Determination of the bond length and binding energy of the Helium
  dimer by diffraction from a transmission grating,
  Phys.\ Rev.\ Lett.\ {\bf 85}, 2284 (2000).

\bibitem{Zel16} 
  S.~Zeller~\textit{et al.},
  Imaging the He$_2$ quantum halo state using a free electron laser,
  Proc.\ Natl.\ Acad.\ Sci.\ U.S.A.\ {\bf 113}, 14651 (2016).

\bibitem{SchToe96}
  W.~Sch\"ollkopf and J.~P.~Toennies,
  The nondestructive detection of the helium dimer and trimer,
  J.\ Chem.\ Phys.\ {\bf 104}, 1155 (1996).

\bibitem{Kun15}
  M. Kunitski~\textit{et al.}, 
  Observation of the Efimov state of the helium trimer,
  Science {\bf 348}, 551 (2015).

\bibitem{BluGre00}
  D.~Blume and C.~H.~Greene,
  Monte Carlo hyperspherical description of helium cluster excited
  states,
  J.\ Chem.\ Phys.\ {\bf 112}, 8053 (2000).

\bibitem{GuaKorNav06}
  R. Guardiola, O. Kornilov, J. Navarro, and J. P. Toennies,
  Magic numbers, excitation levels, and other properties of small neutral
  $^4$He clusters ($N\le50$),
  J. Chem. Phys. {\bf 124}, 084307 (2006).

\bibitem{HiyKam12a}
  E.~Hiyama and M.~Kamimura,
  Variational calculation of $^4$He tetramer ground and excited states
  using a realistic pair potential,
  Phys.\ Rev.\ A {\bf 85}, 022502 (2012).

\bibitem{BazValBar19}
  B. Bazak, M. Valiente, and N. Barnea,
  Universal Short Range Correlations in Bosonic Helium Clusters,
  arXiv:1901.11247.

\bibitem{JanAzi95}
  A.~R.~Janzen and R.~A.~Aziz,
  Modern He--He potentials: Another look at binding energy, effective
  range theory, retardation, and Efimov states,
  J.\ Chem.\ Phys.\ {\bf 103}, 9626 (1995).

\bibitem{PrzCenKom10}
  M.~Przybytek, W.~Cencek, J.~Komasa, G.~Lach, B.~Jeziorski, and K.~Szalewicz,
  Relativistic and quantum electrodynamics effects in the helium pair
  potential,
  Phys.\ Rev.\ Lett.\ {\bf 104}, 183003 (2010).

\bibitem{CenPatSza09}
  W. Cenceka, K. Patkowski, and K. Szalewicz,
  Full-configuration-interaction calculation of three-body nonadditive
  contribution to helium interaction potential,
  J. Chem. Phys. {\bf 131}, 064105 (2009).

\bibitem{ValPhi15}
  M.~Pavón Valderrama and D.~R.~Phillips,
  Power Counting of Contact-Range Currents in Effective Field Theory,
  Phys.\ Rev.\ Lett.\ {\bf 114}, 082502 (2015).

\bibitem{Kirscher:2009aj}
  J. Kirscher, H. W. Grie\ss hammer, D. Shukla, and H.M. Hofmann,
  Universal Correlations in Pion-less EFT with the Resonating Group 
  Model: Three and Four Nucleons,
  Eur.\ Phys.\ J.\ A {\bf 44}, 239 (2010).

\bibitem{Lensky:2016djr}
  V.~Lensky, M. C.~Birse, and N. R.~Walet,
  Description of light nuclei in pionless effective field theory using
  the stochastic variational method,
  Phys.\ Rev.\ C {\bf 94}, 034003 (2016).

\bibitem{Bansal:2017pwn}
  A.~Bansal, S.~Binder, A.~Ekstr\"om, G.~Hagen, G. R.~Jansen, and T.~Papenbrock,
  Pion-less effective field theory for atomic nuclei and lattice nuclei,
  Phys.\ Rev.\ C {\bf 98}, 054301 (2018).

\bibitem{Kievsky:2017mjq} 
  A.~Kievsky, A.~Polls, B.~Juliá-Díaz, and N.~K.~Timofeyuk,
  Saturation properties of helium drops from a Leading Order
  description,
  Phys.\ Rev.\ A {\bf 96}, 040501 (2017).

\end{thebibliography}
\end{document}